\documentclass[12pt]{article}
\usepackage[utf8]{inputenc}
\usepackage{float}
\usepackage{array}
\usepackage[version=4]{mhchem}
\usepackage{graphicx}
\usepackage{tabularx}
\usepackage{booktabs}
\usepackage{amsmath}
\usepackage{changepage}
\usepackage{listings, xcolor}
\usepackage{mathrsfs}
\graphicspath{ {graphics/} }
\usepackage{fancyhdr}
\usepackage{placeins}
\usepackage{titlesec}
\usepackage{listings}
\usepackage{xcolor}

\definecolor{codegreen}{rgb}{0,0.6,0}
\definecolor{codegray}{rgb}{0.5,0.5,0.5}
\definecolor{codepurple}{rgb}{0.58,0,0.82}
\definecolor{backcolour}{rgb}{0.95,0.95,0.92}

\lstdefinestyle{mystyle}{
    backgroundcolor=\color{backcolour},   
    commentstyle=\color{codegreen},
    keywordstyle=\color{magenta},
    numberstyle=\tiny\color{codegray},
    stringstyle=\color{codepurple},
    basicstyle=\ttfamily\footnotesize,
    breakatwhitespace=false,         
    breaklines=true,                 
    captionpos=b,                    
    keepspaces=true,                 
    numbers=left,                    
    numbersep=5pt,                  
    showspaces=false,                
    showstringspaces=false,
    showtabs=false,                  
    tabsize=2
}

\usepackage{caption,tabularx,array,booktabs}
\newcolumntype{C}{>{\centering\arraybackslash}X}
\titleformat{\chapter}{}{}{0em}{\bf\LARGE}
\usepackage[letterpaper,width=150mm,top=25mm,bottom=25mm]{geometry}
\fancypagestyle{plain}{
    \pagestyle{fancy}
    \fancyhead{}
    \fancyhead[LO, LE]{\thepage}
    \fancyhead[RO, RE]{Submission to M3 Challenge 2023 (Honorable Mention)}
    \fancyfoot{}
    \fancyheadoffset{0mm}
}
\begin{document}

\begin{center}
    \LARGE
    \textbf{The Growth of E-Bike Use} \\
    \large
    \textbf{Aditya Gupta, Samarth Chitgopekar, Alexander Kim, Joseph Jiang, Megan Wang, Christopher Grattoni}
\end{center}
\par
\noindent
{\Large \textbf{Executive Summary} \\ \par}
\noindent
\textit{Dear Secretary Buttigieg}, 

We are excited to present our work on foundational questions regarding electric bicycles for policymakers in the United States. In recent years, electric bicycles, or e-bikes, have been increasingly popular \cite{electrek_sales} as people seek fast and eco-friendly methods of transportation. As a result, the rising use of e-bikes is an important factor for policymakers to observe as we shift toward a more sustainable energy plan. We hope our mathematical modeling can help policymakers determine the value of e-bikes and their role in our future.

We first predicted growth in e-bike sales in the U.S. with an ARIMA model, a supervised machine-learning algorithm, that forecasts future e-bike sales based on past sales figures. This model was trained on a dataset of e-bike sales per month from January 2006 to December 2022. To obtain data for U.S. e-bike sales, we used appropriate Lagrange Interpolation on European E-Bike Sales data. Furthermore, we obtained monthly data by using Google Trends data to determine the proportion of annual sales in each month. Based on the training set, our model was able to project e-bike sales two and five years into the future as \textbf{1.3 million} units in 2025 and \textbf{2.113 million} units in 2028. 

Next, we used a Random Forest regression model to determine whether certain factors are significant reasons for the growth of e-bike usage. Such models have shown to work well in other domains i.e game theory \cite{gupta2023determining} \cite{gupta2023value}. To develop this model, we isolated several variables: environmental concern, gas prices, disposable income, and e-bike popularity (quantified by Google Search Trend Data). We then generated the relative importance (model coefficients) of each factor, resulting in the most significant factors in predicting e-bike sale growth being \textbf{disposable personal income}, closely followed by \textbf{popularity}. 

Finally, we quantified the effects of e-bikes on carbon emissions (kilograms/month) and health/wellness (kilocalories burned/month). To quantify these values, we used two separate Monte Carlo simulations \cite{gupta2023using}. First, for both Monte Carlo simulations, we estimated the total number of active e-bikes in the U.S. at a given time using a cumulative sum of the number of bikes sold over the average lifespan of an e-bike. In order to determine the impact of increasing e-bike usage on \ce{CO2} emissions, we used a Monte Carlo simulation that randomly samples from the distributions of an e-bike's expected life, miles ridden on an e-bike, \ce{CO2} emissions from cars, and \ce{CO2} emissions from e-bikes over a 100 times. Our results showed that the total number of kilograms of CO2 reductions from e-bike use in the U.S. in 2022 is $\textbf{15737.82}$ kilograms. By obtaining a sampling distribution for the average number of calories burned per mile driven using an e-bike, we were able to conduct a separate Monte Carlo simulation to predict that the total number of kilo-calories burned from e-bike use in the U.S. in 2022 was $\textbf{716630.727}$ kCal.

\newpage

\tableofcontents

\newpage
\section{Global Assumptions} 
\begin{enumerate}
    \item \textit{Developed countries have a relatively similar rate of growth in e-bike sales.} Developed countries differ from developing countries by economic standing. Therefore, developed countries can be grouped as entities with similar rates of growth.
    \item \textit{Given any month in a specific year, its respective annual proportion of e-bike sales is roughly the same as its respective annual proportion of Google searches.} It can be reasonably assumed that during, or before purchasing an e-bike, consumers will Google the term "electric bikes" and thus, the proportion of annual sales in each month equals the proportion of annual searches in that month. Natural Language Processing has been used in other fields i.e finance \cite{gupta2023analysis}
    \item \textit{Nothing unprecedented will cause future e-bike sales to increase or decrease dramatically.} This includes new types of transportation and other outlying factors, such as pandemics or wars.
    \item \textit{E-bike sales will not be hindered by a lack of e-bike production.} Thus, the rate of e-bike production will not significantly impact sales.
\end{enumerate}
\subsection{Preparing the Data}
The given dataset for e-bike sales lacked consistent historical data. In order to maximize the amount of information we could leverage in our models, we successively applied the following transformations. The code for this procedure is available in \textit{Appendix 5.1}.
\begin{enumerate}
    \item \textit{Merging European data.} {Global Assumption 1} allows us to determine U.S. e-bike sales from 2006 to 2017 using data on European e-bike sales by utilizing a LaGrange interpolation with a first-degree polynomial. For every year from 2006 to 2017 of which we were not provided data on U.S. e-bike sales, we calculated the proportion of European sales that year to European sales in 2019, and used this proportion to find the number of e-bikes sold in the U.S. that year (in relation to U.S. sales in 2019) \cite{mathworks}, \cite{wfsgi}, \cite{mordorintelligence}.
    \item \textit{Computing Monthly Sales.} {Global Assumption 2} allows us to use monthly Google search proportions for the term "electric bikes" \cite{google_trends} in combination with the annual e-bike sale data from above to develop monthly sale data. For a given year, we found the monthly distribution of search frequencies as a proportion relative to the total annual frequency. Then, we found each month's e-bike sales by multiplying its respective search proportion by that year's total annual sales.
\end{enumerate}

\section{Part I: The Road Ahead}

\subsection{Defining the Problem}
In this problem, we were asked to create a model to predict annual e-bike sales in the United States. Using it, we were asked to predict American sales for e-bikes sales two (2025) and five (2028) years from now.

\subsection{Local Assumptions}
\begin{enumerate}
    \item \textit{There exist adequate charging stations in the US transportation system} \cite{bicycle2work}. Given that electric bikes are able to be charged by a standard wall outlet, it can be concluded availability of charging stations does not affect sales.
    \item\textit{The cost of e-bike constituent materials will stay within reasonable bounds.} E-bicycles have a high percent composition of aluminum alloy. The increase in the cost of aluminum alloy is projected to be directly proportional to U.S inflation rates and thus would not pose a significant impact on e-bike production \cite{databridgemarketresearch}. 
    \item \textit{The majority of e-bike owners are adults.} Given that over 40 states in the United States have an age limit or require a license to operate an electric bike\cite{peopleforbikes_state_laws}, it can be concluded that the vast majority of e-bike owners and buyers will be adults. Therefore, potential buyers will not be affected by an inability to operate the vehicle \cite{yougov}.
    \item \textit{Roads and transportation systems will accommodate e-bikes.} The majority of states allow e-bikes to be ridden on roads and bike lanes \cite{peopleforbikes_state_laws}.
    \item \textit{Moving forward, COVID-19 will not pose a significant impact on transportation use.} Thus, future e-bike sales will not be affected by COVID-19.
\end{enumerate}
\subsection{The Model}
\subsubsection{Developing the Model}
In our model, we predicted e-bike sales two and five years into the future.
\vspace{0.4cm}

We used a univariate autoregressive integrated moving average (ARIMA) model to determine future values in a time series based on past values. An ARIMA model is appropriate for predicting time series data and providing an accurate, reliable short-term forecast. Moreover, we have a univariate time series. Thus, a univariate ARIMA model is ideal for this problem. We will use the monthly US e-bike sale data obtained from section 0.1 as our time series data.

In order to use an ARIMA model, we required stationary data, which means that the mean of the series should not be a function of time. Similarly, the series must have the property of homoscedasticity, meaning that the variance of the series remains constant over time \cite{towardsdatascience}. 

The Augmented Dickey-Fuller (ADF) Test is used to determine if the given data is stationary. In the ADF test, 
$$\text{Null Hypothesis, } H_0 : \text{The data is not stationary}$$
$$\text{Alternate Hypothesis, } H_a : \text{The data is stationary}$$

When we performed the test on the raw data, it resulted in a $p$-value of $0.999047$. Since $p > \alpha = 0.05$, this indicated that we failed to reject the null hypothesis (in which the data is not stationary), implying that the data was not yet stationary. To render the data stationary, we took the logarithm of the dependent variable to lower the rate at which the rolling mean changed and differenced the dependent variable to remove effects of time on the units sold(taking the difference between consecutive elements of the dataset). Performing the Augmented Dickey-Fuller Test on the differenced and logarithmic data resulted in a $p$-value of $4.204\cdot10^{-7}$. Since $p < \alpha = 0.05$, we rejected the null hypothesis, meaning that an ARIMA model is appropriate for the logarithmic, order-1 differenced data. Thus, it was very likely the time series was now stationary and that the data's statistical properties were constant over time, so an ARIMA model was further appropriate.

\begin{center}
    \begin{figure}[H]
    \includegraphics[scale = 0.35]{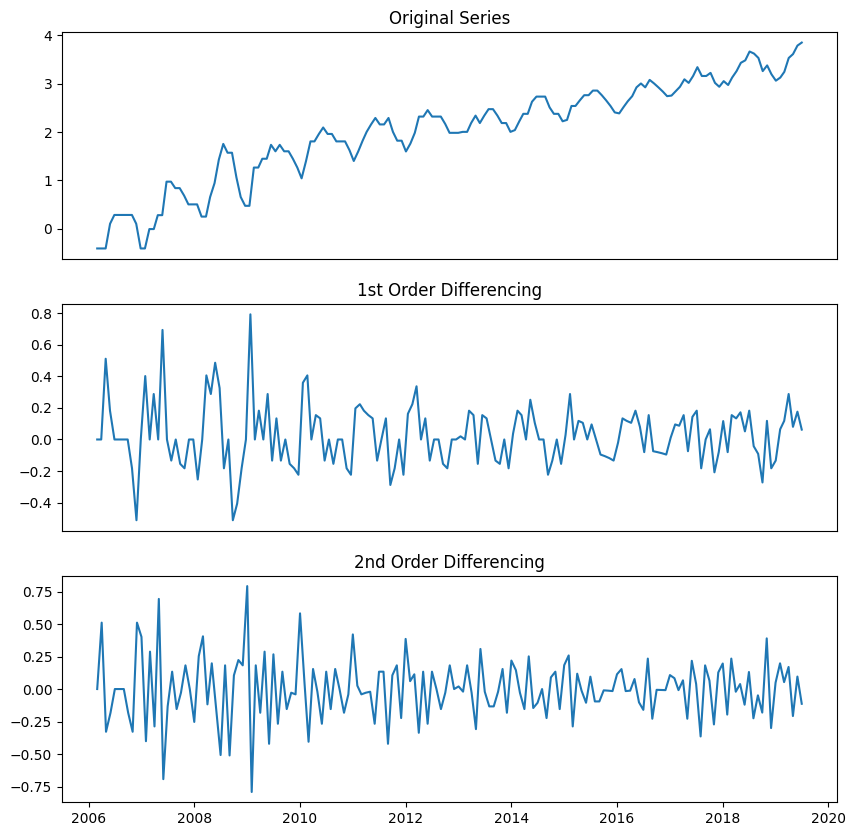}
    \centering
    \caption{}
    \end{figure}
\end{center}

A typical ARIMA model is defined by the integer parameters $(p,d,q)$. $p$ is the number of autoregressive terms, $d$ is the number of differences that need to be taken for stationarity, and $q$ is the number of lagged error terms used to predict the current value \cite{duke_arima}.

Through testing, we determined $p$ to have an optimal value of $12$, meaning that there are $12$ autoregressive terms that the model uses to compute each element of the time series.

We determined $d$ to have a value of $1$ as first-order differencing transformed the time series into one that was stationary and removed seasonal differences. 

Lastly, we gave the model a parameter value of $q=1$, which represents the number of lagged forecast errors that the model uses in its predictions.

\subsubsection{Executing the Model}
\begin{center}
    \begin{figure}[H]
    \includegraphics[scale = 0.45]{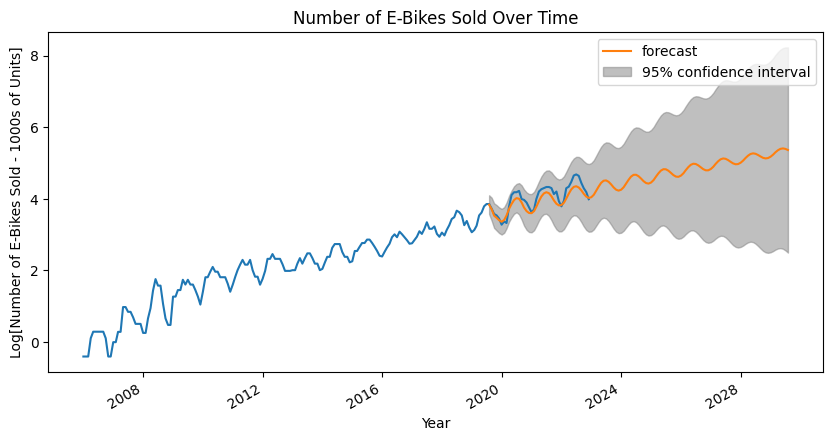}
    \centering
    \caption{}
    \end{figure}
\end{center}
We visualized the data with $\log{(\text{Number of e-bikes sold in 1000s of units)}}$ versus time (in years) graph in order to show the prediction of the model. 

We used the ARIMA Model to predict the next several years of e-bike sales in the U.S., as shown in Figure 2. The complete results of the model are summarized in section 1.5

\subsubsection{Validating the Model}

We used the Ljung-Box test at $\alpha = 0.05$ to determine the overall randomness of the model based on the number of lags (fixed amounts of time change). To run the Ljung-Box test, we used the following calculation, $\displaystyle Q(m) = n (n+2) \sum_{j=1}^{m} \frac{r_j^2} {n - j}$, where $r_j$ is the accumulated sample autocorrelations and $m$ is the time lag. We determined that $Q = {0.90}$. Since $Q = {0.90} > \alpha = 0.05$, we failed to reject the $H_0$ of the test, that the residuals are independently distributed. As a result, we concluded our model does not show a lack of fit. 

\vspace{0.4cm}

To examine the data for outliers, we calculated the kurtosis value. Kurtosis, which can be calculated with $\displaystyle k = \frac{\sum_{i=1}^N (Y_i - \overline{Y})^4} {N \cdot s^4}$ where $\overline{Y}$ is the mean, $s$ is the standard deviation, and $N$ is the number of data points, returned ${0.27}$. Since this value of kurtosis was relatively low $(<3)$, we concluded that the distribution of the data was platykurtic and that the effect of outliers on the data was insignificant. 

\subsection{Strengths and Weaknesses}
Our prediction is strong for short-term forecasts such as two or five years into the future, but, due to the nature of the ARIMA model, is unable to be applied to larger scales of time. The quantity of data we supplied, around 200 data points, was adequate for training an accurate ARIMA model \cite{otexts}. In addition, while the $p, d, \text{and } q$ values used in our model were effective, they were subjectively selected. Moreover, the model did not account for any major turn of events, as noted \textit{Global Assumption 3.}

\subsection{Results}
Our ARIMA model yielded the following forecast.

\begin{center}
    \begin{figure}[H]
    \includegraphics[scale = 0.27]{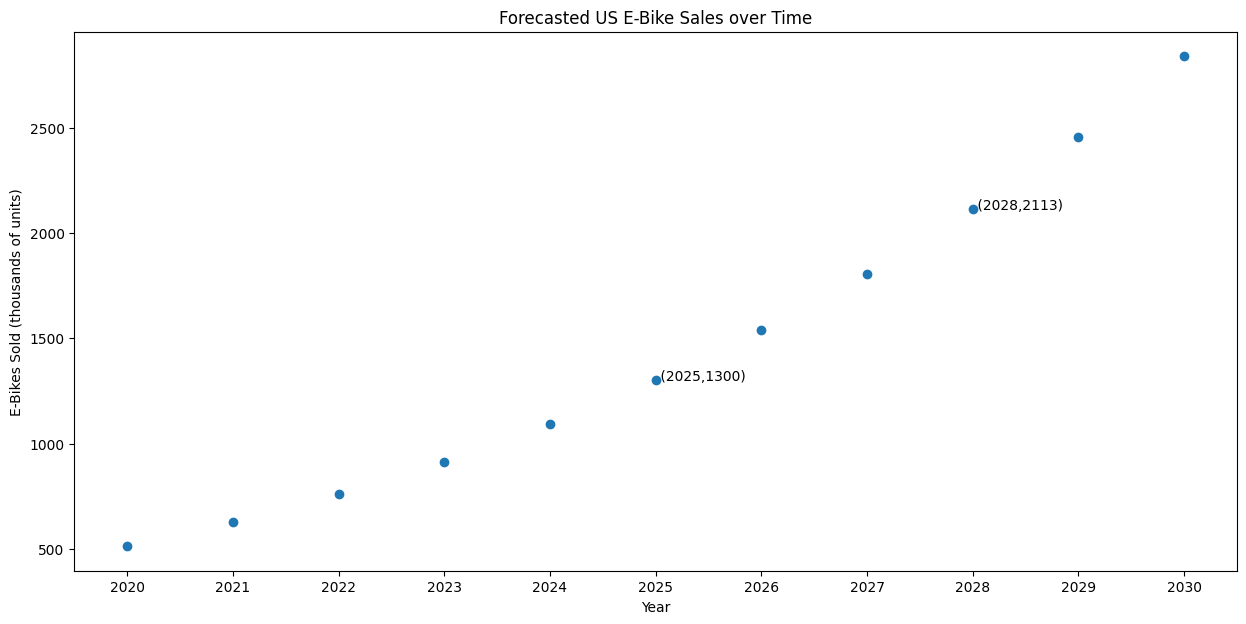}
    \centering
    \caption{}
    \end{figure}
\end{center}

\subsection{Conclusion}
Using our ARIMA model with parameters $(p,d,q)$ of $(12,1,1)$, we predicted that, two years from now, in 2025, there would be $1.300$ million e-bikes sold in the US. Additionally, five years from now, in 2028, there would be $2.113$ million e-bikes sold in the US.

\section{Part II: Shifting Gears}
\subsection{Defining the Problem}
In this problem, we were asked to determine how certain factors affected the growth of e-bike sales in the U.S. and whether the effects of one or more of these factors was significant or not. 
\subsection{Local Assumptions}
\begin{enumerate}
    \item \textit{The popularity ("coolness") factor is directly correlated to Google searches pertaining to electric bikes.} As interest in electric bikes increases, it follows that Google Search frequencies will also increase.
    \item \textit{E-bike usage is only significantly impacted by environmental concerns, gas prices, per capita disposable income in the United States, and popularity.} According to our model, these are the dominant factors that comprise growth in e-bike usage.
    \item \textit{Environmental concern can be extrapolated from those who either worry a great deal or a fair amount about the quality of the environment} \cite{mathworks}. Those who worry only a little or not at all do not contribute meaningfully to environmental concerns.
    \item \textit{Environmental concern does not change significantly throughout individual months of a year.} Environmental concerns do not change significantly throughout the months of the year, as those who worry about the quality of the environment generally do not change their opinion based on the time of year.
    \item \textit{The most common term for e-bikes is "electric bikes".} When comparing the search history of e-bikes, e-bikes, and electric bikes as search terms, electric bikes had the highest proportion of searches.
\end{enumerate}
\subsection{Variables}
The following variables were what we determined to be plausible significant factors in the trend of e-bike usage \cite{mathworks}, \cite{google_trends}. Thus, the main factors we analyzed were disposable personal income, popularity, gas prices, and concern for the environment.

\begin{table}[H]
\resizebox{\textwidth}{!}{%
\begin{tabular}{|c|c|c|c|c|c|}
\hline
\textbf{Year} & \textbf{\begin{tabular}[c]{@{}c@{}}Environmental\\ Concern\end{tabular}} & \textbf{\begin{tabular}[c]{@{}c@{}}Gas Prices in\\  the US (USD)\end{tabular}} & \textbf{\begin{tabular}[c]{@{}c@{}}Per capita disposable\\  Personal Income\\ in the United States, \\ chained (USD)\end{tabular}} & \textbf{\begin{tabular}[c]{@{}c@{}}Google Trends \\ Relative Values \\ Annually\end{tabular}} \\ \hline
2006 & 0.77 & 2.57 & 37570 & 11.583333 \\ \hline
2007 & 0.76 & 2.8 & 38093 & 13.25 \\ \hline
2008 & 0.74 & 3.25 & 38188 & 19.083333 \\ \hline
2009 & 0.71 & 2.35 & 37814 & 13.666667 \\ \hline
2010 & 0.68 & 2.78 & 38282 & 12.583333 \\ \hline
2011 & 0.68 & 3.52 & 38769 & 12.5 \\ \hline
2012 & 0.73 & 3.62 & 39732 & 12 \\ \hline
2013 & 0.69 & 3.51 & 38947 & 11.75 \\ \hline
2014 & 0.66 & 3.36 & 40118 & 12.833333 \\ \hline
2015 & 0.68 & 2.43 & 41383 & 14.083333 \\ \hline
2016 & 0.73 & 2.14 & 41821 & 16.583333 \\ \hline
2017 & 0.77 & 2.42 & 42699 & 20 \\ \hline
2018 & 0.72 & 2.72 & 43886 & 25.416667 \\ \hline
2019 & 0.74 & 2.6 & 44644 & 30.25 \\ \hline
2020 & 0.69 & 2.17 & 47241 & 47.333333 \\ \hline
2021 & 0.75 & 3.01 & 48219 & 59.416667 \\ \hline
2022 & 0.71 & 3.95 & 45405 & 69.916667 \\ \hline
\end{tabular}%
}
\caption{}
\end{table}
\subsection{The Model}
We use a Random Forest algorithm to create a regression model. This model is designed to vary the weighting of the input variables in order to make accurate predictions. A decision tree is a machine-learning process that uses data to make an estimate. Instead of one large decision tree, random forests leverage several smaller trees and take the average of their results to yield more accurate regressions.

For this problem, we will supply the variables defined in Table 1, all of which we suspect could have a role in predicting the sales of e-bikes, and train the regression on the monthly U.S. e-bike sale data from section 0.1.

Then, we can interpret the feature importance of any given variable in the forest as its respective model coefficient. This coefficient tells us how much the incorporation of any given feature reduces the impurity (or improves the accuracy) of the model. Since this is a quantified metric, we are able to leverage it in comparing the significance of various factors in predicting the number of e-bikes sold in the U.S. 

The average prediction error of the model was 1.27 (thousands of units) when tested on the testing data, which was randomly selected from our data set.
\subsubsection{Developing the Model}
We incorporated our variable data as features in a Random Forest, using 70\% of the monthly U.S. e-bike sale dataset from section 0.1 as the training data. The remaining 30\% of the monthly sale dataset is used for testing.

The classification of an individual point in the monthly U.S. e-bike sale dataset as a training or testing point is done randomly until the distribution thresholds have been met so as to avoid any non-uniform bias propagation in the subsets. 

\subsection{Strengths and Weaknesses}
Our model is able to quantify the relative prediction importance across all features. As a result, we are able to go beyond stating whether or not a factor is significant. Instead, we can rank the significance of various features in relation to one another. If two given variables are both arbitrarily deemed significant, our model enables us to comment on which was more important in the regression. However, due to the depth of the decision trees, our model is a black box in the sense that we are unable to explain why the resulting coefficient correlation is the most optimal, even though we know, at a high level, how the model got there. Because the final significance data is accurate, we valued the increased opportunity for comparative analysis and stayed with the model.

\subsection{Results}

After training the Random Forest with the given features, we got the following model coefficient distribution.

\begin{table}[H]
\centering
\resizebox{\textwidth}{!}{%
\begin{tabular}{|c|c|}
\hline
\textbf{Variable}          & \textbf{Computed Model Coefficient} \\ \hline
Disposable Personal Income & 0.48                                \\ \hline
Popularity                 & 0.42                                \\ \hline
Gas Prices                 & 0.08                                \\ \hline
Environmental Concern      & 0.02                                \\ \hline
\end{tabular}%
}
\caption{}
\end{table}
\subsection{Sensitivity Analysis}
The coefficients of the Random Forest are a consolidated sensitivity analysis with relative proportions for each variable in the forest (which is why they sum to 1). As a result, our procedure for analyzing the significance of a given factor is, in and of itself, a sensitivity analysis. The factors we deem the most significant in predicting monthly U.S. e-bike sales (e.g. disposable income, popularity) are those that the model is most sensitive to. Conversely, the factors we deem insignificant (e.g. environmental concern) are those that the model is least sensitive to.

\subsection{Conclusion}
Of the variables analyzed, the most significant in predicting monthly US e-bike sales are disposable personal income and popularity, due to their comparatively large model coefficients (0.48 and 0.42 respectively). These factors make up 90\% of the relative importance across all factors considered.

\section{Part III: Off the Chain}
\subsection{Defining the Problem}
As e-bike usage increases, usage of other methods of transportation decreased. This resulted in an impact on carbon emissions, traffic congestion, health and wellness, and more. In this problem, we were asked to quantify the impact of increased e-bike usage on factors we deemed important.
\subsection{Local Assumptions}
\begin{enumerate}
    \item \textit{Most purchasers of e-bikes will be automobile users.} Those who own conventional bikes likely do so for fitness or recreational factors and thus are unlikely to purchase an e-bike. Consequently, conventional bike users who convert to e-bikes are negligible in comparison to the number of automobile users who convert. 
    \item \textit{The user does not use the bike after its expected life is fulfilled} \cite{gadgetreview}. It is reasonable to assume that once the expected lifecycle of a e-bike is achieved, the user will no longer use this e-bike.
    
    \item \textit{It's assumed that e-bike driving miles are miles that would have been traveled by car.} Any e-bike users likely will utilize their purchase for smaller distances that would normally be traveled by car. Note that this does not imply the inverse, since longer distances will require the use of a motor vehicle.
    \item\textit{The number of e-bikes sold prior to 2006 is negligible in comparison to the number of active e-bikes in use.} This is because the popularity of e-bikes has risen by a significant amount since 2006 and the proportion of e-bikes bought in 2006 is minimal.
    \item\textit{Data pertinent to e-bikes from Vermont is generalized to the whole US}. Vermont has an average temperature comparable to that of the United States \cite{worlddata_vermont_climate}. This similar climate enables us to use data obtained in Vermont for the whole nation. 
    \item \textit{Any e-bikes bought will be used.} E-bikes are expensive enough that consumers investing in one will utilize it.
    \item \textit{The standard deviation of a dataset can be reasonably approximated with the expression}:
    \begin{center}
        $\displaystyle \dfrac{\text{maximum} - \text{minimum}} {4} \text{ }\cite{thoughtco}.$
    \end{center}     
    This is a reasonably accurate approximation given a lack of complete data.
\end{enumerate}
\subsection{Variables}
\begin{table}[H]
\resizebox{\textwidth}{!}{%
\begin{tabular}{|l|l|l|l|l|}

\hline
Variable & Description & Units & Mean & Standard Deviation \\ \hline
E-bike Lifespan ($\text{L}_\text{E}$) & \begin{tabular}[c]{@{}l@{}}Expected e-bike lifespan, after which \\ the user will fully replace the bike, in\\ accordance with \textit{Local Assumption 2}.\end{tabular} & Months & $48_{(23)}$ & 6 \\ \hline
Driving Miles (d) & \begin{tabular}[c]{@{}l@{}}Average miles ridden per month by e-bike users that, by\\ \textit{Local Assumption 3}, would otherwise\\ have been traveled by car.\end{tabular} & Miles/Month & $63.33_{(12)}$ & 12.02 \\ \hline
\begin{tabular}[c]{@{}l@{}}Car \ce{CO_2} \\ Emissions ($\text{E}_{\text{car}}$)\end{tabular} & \begin{tabular}[c]{@{}l@{}}The average carbon dioxide emitted\\ from a car per mile driven\end{tabular} & g/Mile & $431.2_{(13)}$ & $107.803_{(13)}$ \\ \hline
\begin{tabular}[c]{@{}l@{}}E-bike \ce{CO_2}\\ Emissions ($\text{E}_{\text{bike}}$) \end{tabular} & \begin{tabular}[c]{@{}l@{}}The average carbon dioxide emitted\\ from an e-bike per mile driven\end{tabular} & g/Mile & $9.01_{(24)}$ & 1.9308 \\ \hline
\begin{tabular}[c]{@{}l@{}}Calories\\ Burned Per Mile (C)\end{tabular} & \begin{tabular}[c]{@{}l@{}}The amount of calories burned each\\ mile spent riding an e-bike\end{tabular} & cals/Mile & 21 & 3.04 \\ \hline
\end{tabular}%
}
\caption{}
\end{table}

\begin{center}
    \textit{Note: Bracketed numbers in the table refer to citations.}
\end{center}

To calculate the value for calories burned per mile, we used the information from sources 25 and 26. From source 25, we learned that the calories burned on electric bikes range from $358 \text{ cal}/\text{hr}$ to $650 \text{ cal}/\text{hr}$. In addition, source 26 stated that e-bike users bike approximately $24 \text{ miles}/\text{hr}$. We used the average calories burned on electric bikes, which computed to $504 \text{ cal} / \text{hr}$. To find the average calories burned per mile, we used the following conversion:

\begin{center}
 $\displaystyle \frac{504 \text{ cal}} {1 {\text{ hr}}} \times \frac{1 \text{ hr}} {24 \text{ miles}} = 21 \text{ cal} / \text{mile}$
\end{center}

Furthermore, the standard deviations lacking citations were calculated based on \textit{Local Assumption 7}.

\subsection{The Model}
We first used a Monte Carlo simulation to predict the effect of the increasing use of e-bikes on \ce{CO2} emissions.

We simulated 100 times the expected decrease, in kilograms, of \ce{CO2} emissions. To create our model, we first found the kilograms \ce{CO2} saved per mile driven with an electric bike instead of a car, yielding the following:
$$\text{F}_{\text{car}} - \text{F}_{\text{bike}}$$
We next found the number of estimated total working bikes in the U.S. at any given point as the cumulative sum of the number of bikes sold in the last $L_e$ months, calling this variable $T_b$. Next, we multiplied the total number of bikes we have by the average number of miles ridden on an e-bike by e-bike owners in a month, $d$ to find the total amount of \ce{CO2} kilograms saved every month. Thus, to find the total amount of \ce{CO2} kilograms saved every months, $TC_m$, we used the following equation: 
$$TC_m = \frac{(\text{F}_{\text{car}} - \text{F}_{\text{bike}}) \cdot T_b \cdot d}{1000}$$

By using a Monte Carlo simulation, we were able to approximate the value of each parameter by randomly sampling a value for the parameter in the normal distribution using the Mean and Standard Deviation given in Table 3. 

We next simulated 100 times the expected number of total kilo-calories burned per month from e-bike use in the U.S. We again used $T_b \cdot d$ to estimate the total number of miles driven by e-bikes. We then multiplied this value by the average number of calories burnt per mile riding e-bikes, giving us the total number of kilo-calories burned per month from e-bike use in the U.S., $TCal_m$ as 
$$TCal_m =\frac{C \cdot T_b \cdot d}{1000}$$

By using a Monte Carlo simulation, we were once again able to approximate the value of each parameter by randomly sampling a value for the parameter in the normal distribution using the Mean and Standard Deviation given in Table 3. 

\subsection{Strengths and Weaknesses}
The Monte Carlo simulation does not take into account financial crises such as the 2008 recession and other unforeseen events. And, possible errors in our standard deviation calculations and means may diminish the accuracy of the model. However, it provides a good basis for quantifying the impacts of e-bikes on various factors. It allows us to reduce uncertainty by changing various parameters and viewing the impact of the factors we established.

\subsection{Results}

We simulate the kilograms of \ce{CO2} reductions every month in the U.S. 100 times, giving us the following figure. 

\begin{center}
    \begin{figure}[H]
    \includegraphics[scale = 0.65]{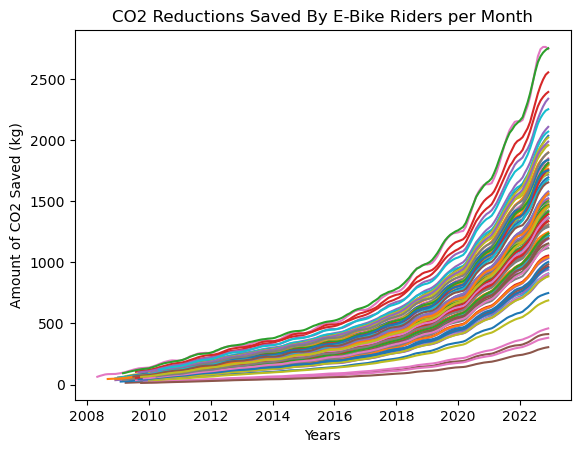}
    \centering
    \caption{}
    \end{figure}
\end{center}

We then simulated the total number of kilo-calories burned per month from e-bike use in the U.S. 100 times, giving us the following figure. 

\begin{center}
    \begin{figure}[H]
    \includegraphics[scale = 0.65]{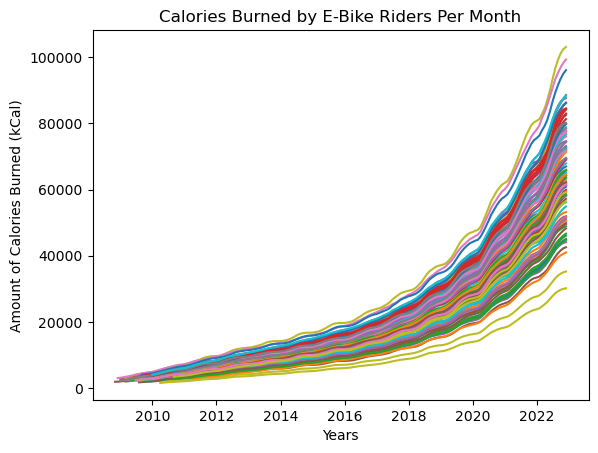}
    \centering
    \caption{}
    \end{figure}
\end{center}

\begin{center}
    \begin{figure}[H]
    \includegraphics[scale = 0.75]{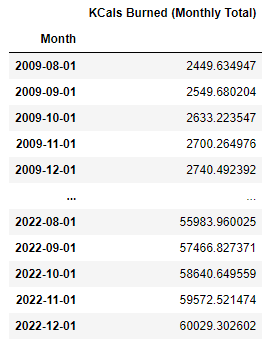}
    \centering
    \caption{}
    \end{figure}
\end{center}

Figure 6 shows one sample simulation generated by the Monte Carlo model of the kilocalories burned due to e-bikes in the U.S. in 2009 and 2022. 

\begin{center}
    \begin{figure}[H]
    \includegraphics[scale = 0.75]{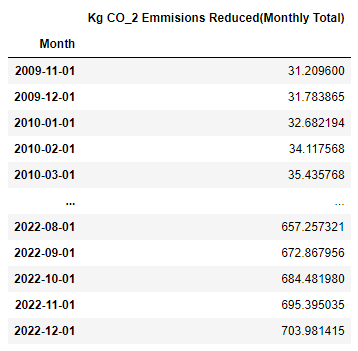}
    \centering
    \caption{}
    \end{figure}
\end{center}

Figure 7 shows one sample simulation generated by the Monte Carlo model of the kilograms of \ce{CO2} emissions reduced due to e-bikes in the U.S. in 2009 and 2022.

\subsection{Sensitivity Analysis}
Since a Monte Carlo simulation accounts for variance in original inputs by randomly selecting from a given distribution, the simulation in and of itself already takes into consideration the possible variation in input variables. Consequently, further sensitivity analysis is not necessary or useful.

\subsection{Conclusion}
Based on our Monte Carlo model, we determined the quantified impact of growth in e-bike usage. The growth resulted in an average decrease of \textbf{15737.82} kilograms of \ce{CO2} emissions in 2022. Moreover, with regards to wellness, increased e-bike usage resulted in a total of \textbf{716630.727} kilo-calories burned in 2022. 
\newpage

\bibliographystyle{plain}
\bibliography{ref}

\begin{thebibliography}{10}
\bibitem{electrek_sales}
Electric bicycles are now outselling electric cars and plug-in hybrids combined
  in the us.
\end{thebibliography}

\newpage

\section{Appendix}

\subsection{Preparing the Data}
\begin{lstlisting}[language=Python]
import csv

lookup = {}
total_freq = 0

with open('../data/ElectricBikesGoogleSearchTrends.csv', 'r') as f:
    # CSV has shape YYYY-MM, relative Google search frequency for electric bikes (0 - 100)
    reader = csv.reader(f.readlines())
    for i, row in enumerate(reader):
        if i == 0 or not len(row): continue
        year = row[0].split('-')[0]
        month = row[0].split('-')\cite{mathworks}
        if not year in lookup:
            lookup[year] = []
        lookup[year].append(int(row\cite{mathworks}))

def get_us_data(european_sales_annual, european_sales_2019, us_sales_2019):
    "Gets US Data from European Data"
    return us_sales_2019*(european_sales_annual/european_sales_2019)

text = 'Month,Sales'
with open('../data/AnnualUSEbikeSales.csv', 'r') as f:
    # CSV has shape year, US sales in thousands of units
    reader = csv.reader(f.readlines())

    for i, row in enumerate(reader):
        if i == 0: continue
        year = row[0]
        total_sales = float(row\cite{mathworks})
        total_frequency = sum(lookup[year])
        for j, monthly_frequency in enumerate(lookup[year]):
            prop = monthly_frequency/total_frequency
            month = str(j + 1).rjust(2, '0')
            text += f'\n{year}-{month},{prop * total_sales}'

with open('../data/MonthlyUSEbikeSales.csv', 'w') as f:
    # CSV has shape YYYY-MM, US sales in month in thousands of units
    f.write(text)
\end{lstlisting}

\newpage

\subsection{Part I}
\begin{lstlisting}[language=Python]
import numpy as np
import pandas as pd
import csv
from matplotlib import pyplot as plt
from statsmodels.tsa.stattools import adfuller
from statsmodels.tsa.seasonal import seasonal_decompose
from statsmodels.tsa.arima.model import ARIMA
from statsmodels.graphics.tsaplots import plot_acf, plot_pacf, plot_predict
import statsmodels.api as sm
import pmdarima as pm

from pandas.plotting import register_matplotlib_converters
register_matplotlib_converters()

df = pd.read_csv('../data/MonthlyUSEbikeSales.csv', parse_dates = ['Month'], index_col = ['Month'])
df = np.log(df)
df.head()
plt.xlabel('Month')
plt.ylabel('Number of Ebikes Sold - 1000s of Units')
plt.plot(df)

plt.figure(figsize=(15,7))
plt.plot(df["US Ebike Sales"], label='Original')
plt.plot(df["US Ebike Sales"].rolling(window=3).mean(), color='red', label='Rolling mean')
plt.plot(df["US Ebike Sales"].rolling(window=3).std(), color='green', label='Rolling std')
plt.xlabel('Year', fontsize=12)
plt.ylabel('US Ebike Sales', fontsize=12)
plt.legend(loc='best')
plt.title('Rolling Mean & Standard Deviation')

def ADF_test(timeseries):
    print("Results of Dickey-Fuller Test:")
    dftest = adfuller(timeseries, autolag="AIC")
    dfoutput = pd.Series(
        dftest[0:4],
        index=[
            "Test Statistic",
            "p-value",
            "Lags Used",
            "Number of Observations Used",
        ],
    )
    for key, value in dftest\cite{google_trends}.items():
        dfoutput["Critical Value (%s)" % key] = value
    print(dfoutput)

ADF_test(df)

train_data = df['US Ebike Sales'][:round(int(len(df)*0.8))]
test_data = df['US Ebike Sales'][round(int(len(df)*0.2)):]

# Original Series
fig, (ax1, ax2, ax3) = plt.subplots(3, figsize=(10, 10))
ax1.plot(train_data)
ax1.set_title('Original Series')
ax1.axes.xaxis.set_visible(False)
# 1st Differencing
ax2.plot(train_data.diff())
ax2.set_title('1st Order Differencing')
ax2.axes.xaxis.set_visible(False)
# 2nd Differencing
ax3.plot(train_data.diff().diff())
ax3.set_title('2nd Order Differencing')
plt.show()

ARIMA_model = pm.auto_arima(train_data, m=3)
ARIMA_model.summary()

model = ARIMA(train_data, order=(12, 1, 1))
model_fit = model.fit()
model_fit.summary()

fig, axs = plt.subplots(1, 1, figsize=(10, 5))
plt.plot(df)
plot_predict(model_fit, start=len(train_data), end=len(train_data) + 12*10, ax=axs)
plt.xlabel('Year')
plt.ylabel('Log[Number of E-bikes sold in thousands of units]')
plt.show()

monthly_forecast = np.exp(model_fit.forecast(137))

df = monthly_forecast.to_frame().to_csv()
with open('../data/MonthlyEBikeUSForecast.csv', 'w') as f:
    f.write(df)

prediction = {}
with open('../data/MonthlyEBikeUSForecast.csv', 'r') as f:
    rows = csv.reader(f.readlines(), delimiter=',')
    for row in rows:
        year = row[0].split('-')[0]
        if not year in prediction: prediction[year] = 0
        prediction[year] += float(row\cite{mathworks})

x = list(prediction.keys())
y = list(prediction.values())
plt.title("US E-Bike Sales (Forecast)")
plt.xlabel("Year")
plt.ylabel("E-Bike Sales (thousands of units)")
plt.scatter(x,y)
plt.text('2025', y[x.index('2025')], f' (2025,{round(y[x.index("2025")])})')
plt.text('2028', y[x.index('2028')], f' (2028,{round(y[x.index("2028")])})')
plt.show()
\end{lstlisting}

\subsection{Part II}
\begin{lstlisting}[language=Python]
import pandas as pd
from sklearn.model_selection import train_test_split
from sklearn.preprocessing import StandardScaler
from sklearn import datasets
import numpy as np

df = pd.read_csv('../data/factors_annual.csv', parse_dates = ['Year'], index_col = ['Year'])
print(df.head)
print(df.describe())

X_train, X_test, y_train, y_test = train_test_split(df.iloc[:, :-1], df.iloc[:, -1:], test_size = 0.3, random_state=1)

cols = ["Environmental Awareness", "Gas Prices in the US (USD)","Disposable Personal Income", "Google Trends Relative Values Annually"]

sc = StandardScaler()
sc.fit(X_train)
X_train_std = sc.transform(X_train)
X_test_std = sc.transform(X_test)

from sklearn.ensemble import RandomForestRegressor

# Instantiate model with 1000 decision trees
rf = RandomForestRegressor(n_estimators = 1000, random_state = 42)

# Train the model on training data
rf.fit(X_train, y_train.values.ravel());

# Use the forest's predict method on the test data
predictions = rf.predict(X_test)

# Calculate the absolute errors
errors = abs(predictions - y_test.values)

# Print out the mean absolute error
print('Mean Absolute Error:', round(np.mean(errors), 2), 'Units (In Thousands)')

# Calculate mean absolute percentage error (MAPE)
mape = 100 * (errors / y_test.values)
print(mape)

# Calculate and display accuracy
print(np.mean(mape))

accuracy = 100 - np.mean(mape)
print('Accuracy:', round(accuracy, 2), '%.')

# Get numerical feature importances
importances = list(rf.feature_importances_)

# List of tuples with variable and importance
feature_importances = [(feature, round(importance, 2)) for feature, importance in zip(cols, importance)]

# Sort the feature importances by most important first
feature_importances = sorted(feature_importances, key = lambda x: x\cite{mathworks}, reverse = True)

# Print out the feature and importance 
[print('Variable: {:20} Importance: {}'.format(*pair)) for pair in feature_importances];

\end{lstlisting}

\subsection{Part III}
\begin{lstlisting}[language=Python]
import pandas as pd
import numpy as np
import random as rand
import matplotlib.pyplot as plt

monthly_bike_sales = pd.read_csv('../data/MonthlyUSEBikeSales.csv', parse_dates = ['Month'], index_col = ['Month'])
monthly_bike_sales

def get_std(min, max):
    return (max - min) / 4

def simulate_c2(dataframe, expected_life_mean, expected_life_std, driving_miles_mean, driving_miles_std, co2_per_car_mean, co2_per_car_std, co2_saved_bike_mean, co2_saved_bike_std):
    data = dataframe.copy(deep = True)
    expected_life = int(np.random.normal(expected_life_mean, expected_life_std))
    driving_miles = np.random.normal(driving_miles_mean, driving_miles_std)
    co2_per_car = np.random.normal(co2_per_car_mean, co2_per_car_std)
    co2_per_bike = np.random.normal(co2_saved_bike_mean, co2_saved_bike_std)
    co2_saved = co2_per_car - co2_per_bike
    data['Estimated Total Working Bikes'] = data['US Ebike Sales'].rolling(window = expected_life).mean()
    data.dropna(inplace = True)
    data['Driving Miles Saved'] = data['Estimated Total Working Bikes'] * driving_miles
    data['CO2 Saved'] = data['Driving Miles Saved'] * co2_saved
    return data

co2_2023_saved = []
for i in range(100):
    dt = simulate_c2(dataframe = monthly_bike_sales, expected_life_mean = 48, expected_life_std = 12 * get_std(3, 5), 
                       driving_miles_mean = 760, driving_miles_std = get_std(580, 1157), co2_per_car_mean = 268 * 1.609, co2_per_car_std = 67 * 1.609, co2_saved_bike_mean = 5.6 * 1.609, co2_saved_bike_std = get_std(3.2, 8) * 1.609)
    plt.plot(dt['CO2 Saved'])
    dt = dt.rolling(12).sum()
    (co2_2023_saved).append(dt['CO2 Saved'].iloc[-1])
plt.xlabel("Years")
plt.ylabel("Amount")
plt.title(("CO2 Saved"))
plt.show()

print('Average CO2 in kilograms saved in 2023', np.mean(co2_2023_saved)/1000)

def simulate_cals_burned(dataframe, expected_life_mean, expected_life_std, driving_miles_mean, driving_miles_std, cals_per_mile_mean, cals_per_mile_std):
    data = dataframe.copy(deep = True)
    expected_life = int(np.random.normal(expected_life_mean, expected_life_std))
    driving_miles = np.random.normal(driving_miles_mean, driving_miles_std)
    cals_per_mile = np.random.normal(cals_per_mile_mean, cals_per_mile_std)
    data['Estimated Total Working Bikes'] = data['US Ebike Sales'].rolling(window = expected_life).mean()
    data.dropna(inplace = True)
    data['Driving Miles Saved'] = data['Estimated Total Working Bikes'] * driving_miles
    data['kCals Burned'] = data['Driving Miles Saved'] * cals_per_mile
    return data

cals_burned_2023 = []
for i in range(100):
    dt = simulate_cals_burned(dataframe = monthly_bike_sales, expected_life_mean = 48, expected_life_std = 12 * get_std(3, 5), 
                       driving_miles_mean = 760, driving_miles_std = get_std(580, 1157), cals_per_mile_mean = 18.58, cals_per_mile_std = get_std(12.5, 16.66))
    plt.plot(dt['kCals Burned'])
    dt = dt.rolling(12).sum()
    (cals_burned_2023).append(dt['kCals Burned'].iloc[-1])
plt.xlabel("Years")
plt.ylabel("Amount")
plt.title(("Kilocalories Burned"))
plt.show()

print('Expected kilocalories Burned by all E-bikes in 2023', np.mean(cals_burned_2023))

\end{lstlisting}

\end{document}